\documentclass{elsart}
\usepackage{graphicx,harvard,amssymb}

\makeatletter
\def\elsartstyle{%
	\def\normalsize{\@setfontsize\normalsize\@xiipt{14.5}}
	\def\small{\@setfontsize\small\@xipt{13.6}}
	\let\footnotesize=\small
	\def\large{\@setfontsize\large\@xivpt{18}}
	\def\Large{\@setfontsize\Large\@xviipt{22}}
	\skip\@mpfootins = 18\p@ \@plus 2\p@
	\normalsize
}
\makeatother

\def\url#1{{\ttfamily\def\/{/\discretionary{}{}{}}#1}}

\begin{document}
\begin{frontmatter} 
\title{Non linear predictions from linear theories in models with Dark Energy}

\author{R. Mainini, A.V. Macci\`o \& S.A. Bonometto}
\address{Physics Department G. Occhialini, Universit\`a degli Studi di
Milano--Bicocca, Piazza della Scienza 3, I20126 Milano (Italy)}
\address{I.N.F.N., Via Celoria 16, I20133 Milano (Italy)}

\thanks[email]{E-mails: mainini@mib.infn.it, maccio@mib.infn.it, 
bonometto@mib.infn.it}

\begin{abstract}
We study the cluster mass function and its evolution in different models
with Dark Energy arising from a self--interacting scalar field, with 
Ratra--Peebles and SUGRA potentials. Computations are based on a 
Press \& Schechter approximation. The mass functions we obtain
are compared with results holding for open models or models with
Dark Energy due to a cosmological constant. Evolution results, in
some Dark Energy models, closely approach open models.
\end{abstract}


\end{frontmatter}

\section{Introduction}
One of the main puzzles in modern cosmology is the nature of Dark 
Energy (DE), whose presence seems to be required by SNIa data 
(see, e.g., Perlmutter et al 1999, Riess et al 1998). 
A joint analysis of CMB and LSS observations (see, e.g., 
Percival et al. 2002, Efstathiou et al 2002) also
favor a flat Universe with a matter density parameter 
$\Omega_m \simeq 0.3$, mostly due to CDM and with a minor 
contribution of baryons ($\Omega_b h^2 \simeq 0.02$;
$h$ is the Hubble constant in units of 100 km/s/Mpc; in this paper
we shall take $h = 0.7$ and, unless differently specified, 
$\Omega_m = 0.3$, anywhere). The residual energy content of the 
world, in the present epoch, should not be 
observable in the number--of--particle representation.

One of the most appealing possibilities is that such dark component
arises from a self--interacting scalar field. With in the wide set
of interaction potential suggested, a particular relevance is kept
by Ratra--Peebles (1988, RP hereafter; see also Wetterich 1995) and SUGRA 
(Brax \& Martin 1999, 2000) expressions:
$$
V(\phi) = \Lambda^{4+\alpha}/\phi^\alpha
~,~~~~~~~~~
V(\phi) = (\Lambda^{4+\alpha}/\phi^\alpha) \exp (\kappa \phi^2/2)~.
\eqno (1)
$$
Here $\Lambda$ is an energy scale, currently set in the range
$10^2$--$10^{10}\, $GeV, relevant for fundamental interaction physics; 
potentials depend also on the exponent $\alpha$; fixing
$\Lambda$ and $\alpha$, the DE density parameter $\Omega_{DE}$
is automatically fixed; in this work we preferred to use as
free parameters $\Lambda$ and $\Omega_{DE}$;  in 
SUGRA potentials, $\kappa = 8\pi G$ ($G$: gravitational constant).

In this work we try to determine some effects on galaxy clusters 
and their evolution, caused by replacing a simple
cosmological constant with DE due to a scalar field
self--interacting  according to the potentials (1).

The technique used to study non--linear evolution is the Press 
\& Schechter (PS, Press \& Schechter 1974) approach. It is based 
on the spherical collapse model, that Gunn \& Gott (1972), Gott \& 
Rees (1975), Peebles (1980) debated within the frame of pure 
CDM models, and Lahav et al (1991), Eke et al (1996), Brian \&
Norman (1998) and others generalized to the case of $\Lambda $CDM. 
In spite of its approximation, such model, inserted in PS formulation, 
has been found to approach simulation results (see, e.g.,  
Lacey \& Cole 1993, 1994). Recent improvements of the method
(Sheth \& Tormen 1999, 2002, Sheth, Mo \& Tormen, 2001, see also
Jenkins et al 2001), allowing a better approximation, involve some 
more parameters and their use seems unnecessary when aiming just to 
compare different cosmological models.

\section{From linear theory to non linear predictions}
In order to apply the PS technique we first need to determine
the amplitude $\delta_c$, in the linear theory, of a spherical fluctuation
that would achieve full recollapse at a given redshift $z_{col}$.
Real fluctuations, after achieving maximum expansion, turn around
and begin to recontract. However, contraction requires that potential 
energy, turned into internal kinetic energy, is radiated soon. If 
radiation is negligible, contraction is delayed and, at $z_{col}$, 
virial equilibrium is attained. In a standard CDM model, $\delta_c$ is 
constant and holds $\sim 1.68$ (see, e.g., Coles \& Lucchin 1995). 
In RP or SUGRA models, $\delta_c$ depends upon the cosmological 
parameters and the redshift $z_{col}$. In Fig.s~1 and 2 we report 
such dependence. As a by--product of such computation, we obtain the 
density contrast ($\Delta_c$) of a spherical fluctuation when
it is fully virialized. In Fig.~3 we report the dependence of
$\Delta_c$ on $\Omega_m$ at $z=0$, for a number of DE models.
Then, in Fig.~4, we report how $\Delta_c$ depends on $z$ 
for some DE models, keeping $\Omega_m = 0.3$ and $h=0.7$.

\begin{figure}
\leftmargin=-0.5pc
\begin{center}
\includegraphics*[width=9cm]{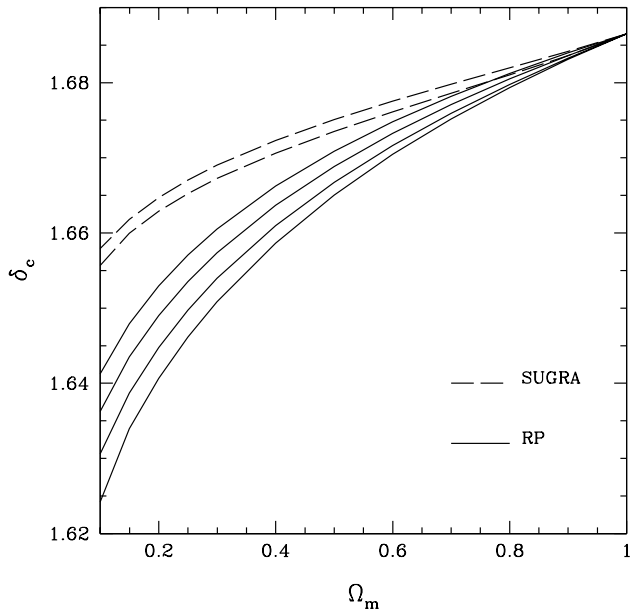}
\end{center}
\caption{The dependence of $\delta_c$ on the matter density parameter
$\Omega_m$ is shown, for 4 RP ($\Lambda/$GeV =  $10^2$, $10^4$,
$10^6$ and 10$^8$) and 2 SUGRA models ($\Lambda/$GeV =  $10^2$ and
$10^8$). $\Lambda$ values increase from top to bottom curves.
}
\end{figure}

\begin{figure}
\begin{center}
\includegraphics*[width=9cm]{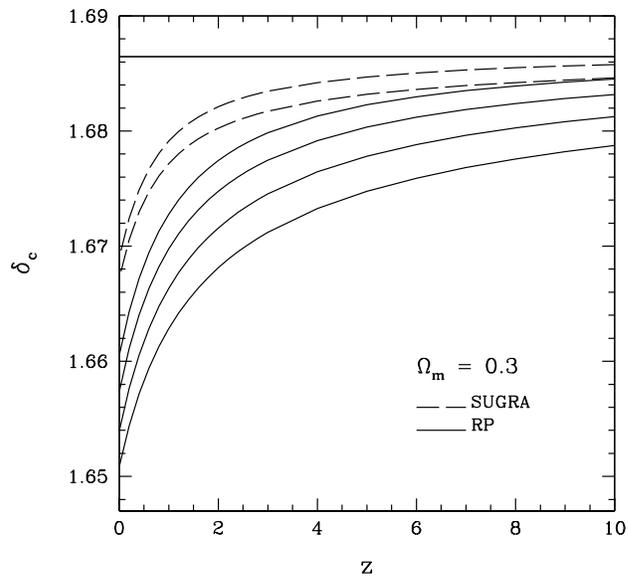}
\end{center}
\caption{The dependence of $\delta_c$ on the redshift $z$ is shown
for $\Omega_m = 0.3$ and $h=0.7$.
$\Lambda$ values and curve setting are the same as in Fig.~1.
}
\end{figure}

The value of $\delta_c$ is to be used in the expression
$$
f(\nu)\nu \, d\log \nu = {M \over \bar \rho_m} n_c(M) M\, d\log M~;
\eqno (2)
$$
(here $\bar \rho_m$ is background matter density, 
the bias factor $\nu = \delta_c/\sigma_M$, $\sigma_M$ 
being the rms density fluctuation on the length scale 
corresponding to the mass $M$), yielding the differential mass function
$n_c(M)$ once the distribution on bias is given.
Here, as usual, we assume a Gaussian $f(\nu)$.
Eq.~(2) shall then be integrated, to obtain $n_c(>M)$.
Let us outline that, in order to compute $\sigma_M$, we need to
know the transfered spectrum $P(k)$; let us also recall the relation
$k = 2\pi (4\pi {\bar \rho}/3 M)^{1/3}$ between wavenumber and mass.

The first finding of this analysis is that the difference between the
mass functions for RP or SUGRA models and the mass function of a 
$\Lambda$CDM model, at $z=0$, is smaller than the difference due to
a shift by 0.05 of the primeval spectral index. The intrinsic 
noise of data is then a serious obstacle to any attempt to determine 
the nature of DE using cumulative mass functions at $z=0$.

On the contrary, we found a clear imprint of the nature of DE 
in the evolution of the number $N$ of clusters (above a suitable
mass $M$, in a box of side $s$). In Fig.~5, we give $N$ (for
$M > 6.9 \cdot 10^{14}h^{-1} M_\odot$ and $s=100\, h^{-1}$Mpc)
as a function of redshift, normalized to an identical number
of clusters at $z=0$, for all models ($N(z=0) = 0.13$).
In Fig.~6 we give the ratio between the
number of clusters expected in various models and the number 
expected in an open CDM model with the same value of $\Omega_m$.
The mass here is selected so to correspond to a cluster with
Abell radius in a standard CDM model. A similar plot, for 
a slightly smaller mass, was given by Bahcall, Fan \& Cen (1997),
for standard CDM, $\Lambda$CDM and OCDM only (for a recent analysis
of the constraints that cluster number counts set to the cosmological
model, see, e.g., Holder, Haiman \& Mohr, 2001).

\begin{figure}
\begin{center}
\includegraphics*[width=8.5cm]{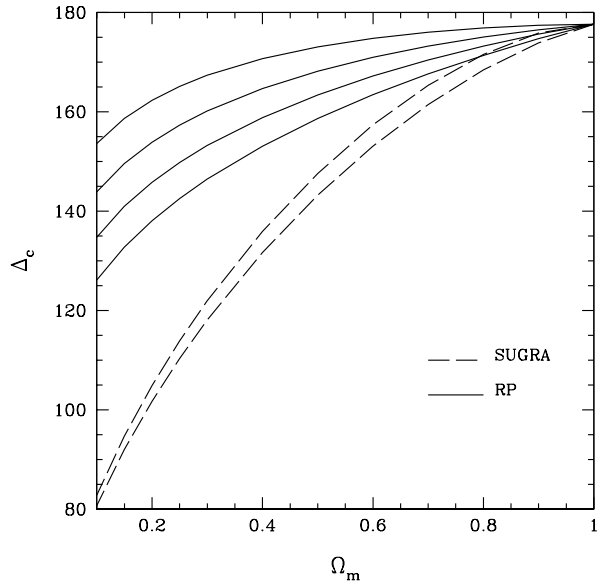}
\end{center}
\caption{The density contrast $\Delta_c$ of a virialized sphere is shown
for 4 RP and 2 SUGRA models as a function of $\Omega_m$,
for $h=0.7$, at $z=0$. 
$\Lambda $ values are the same as in Fig.~1, but here they
decrease from top to bottom curves.
}
\end{figure}

\begin{figure}
\begin{center}
\includegraphics*[width=8.5cm]{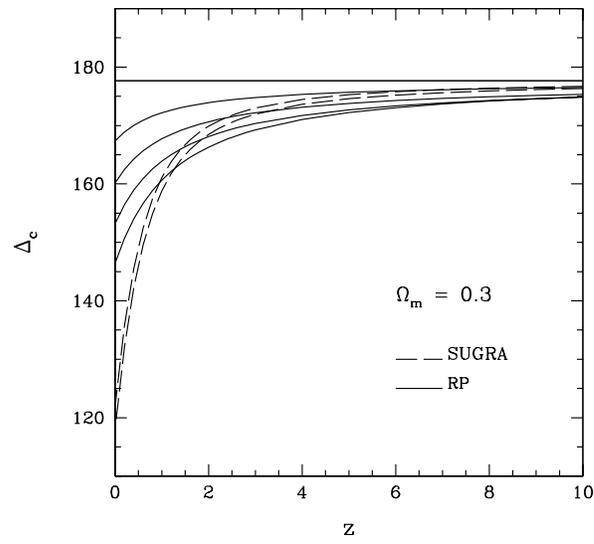}
\end{center}
\caption{The density contrast $\Delta_c$ of a virialized sphere is shown
for 4 RP and 2 SUGRA models as a function of $z$. For the sake of 
comparison, the constant value of $\Delta_c$ in standard CDM is also 
reported. $\Lambda $ values and curve setting are the same as in Fig.~3.
}
\end{figure}

\section{Discussion}
The above mass functions and evolution are calculated using a 
transfer function obtained from a generalization of the public 
program CMBFAST to cosmologies with DE given by the potentials 
(1). Initial conditions were set according to the tracker solution
in radiation dominated era (Steinhardt, Wang \& Zlatev 1999;
Zlatev, Wang, \& Steinhardt 1999, Brax, Martin \& Riazuelo 2000). 
DE fluctuations, taken into account in primeval fluctuation evolution,
are damped soon after a scale enters the horizon and, therefore, 
are not important in further fluctuation evolution. Accordingly, 
in our problem, their relevance amounts to causing some modification 
of the transfer function. (On the contrary, they cause 
major modifications of CMB anisotropy and polarization.)

Fig.~7 describes the pattern followed to evaluate $\delta_c$ and
$\Delta_c$ for $z_{col} = 0$; for $z_{col} > 0$ the procedure is 
similar. It must be however outlined that Fig.~2 does not describe
a time evolution, but is worked out considering different time
evolutions of the kind described in Fig.~7. We obtain the linear 
growth factor by using a simplified program, suitable to treat 
fluctuation evolution after recombination, which includes the 
Friedman equation, the equation of motion of the scalar field and 
the Jeans' equation for the linear density fluctuation $\delta$:
$$
\ddot \delta + 2{\dot a \over a} \dot \delta - {3 \over 2} {H_o \Omega_m
\over a^3} \delta = 0
\eqno (3)
$$
(dots yield derivatives in respect to time, $H_o$ is the present 
value of the Hubble constant). Such system provides the lower curves
in Fig.~7; for a flat model without DE (SCDM), the curve is a straight 
line; together with it, in the figure, we show the behavior for a RP model. 
Other models, with different DE contents, have a similar behavior. 
Such linear results can 
be multiplied by an arbitrary factor and, in the figure, they are 
normalized so that the initial fluctuation amplitude allows full 
recollapse, at $z_{col}$, of a spherically symmetric fluctuation. 
This is also to be done in actual calculations to evaluate $\delta_c$. 
The evolution of the contrasts $\Delta=\rho_m/{\bar \rho_m}$, between
inner and average matter densities, is then shown by the upper curves. 
However, instead of reporting their diverging at $z_{col}$, we interrupt 
their growth when virialization density contrast is attained;
the residual slight growth of $\Delta$ is due to cosmic expansion.
Notice that the density contrasts $\Delta_c$, shown in Fig.s 3 and 4,
are $\Omega_m \Delta$.

More in detail, within such spherical lump, at an initial redshift $z_i$, 
the energy density exceeds average by a factor $1+\delta_i = \Delta_i$,
tuned so cause full recollapse at an assigned redshift $z_{col} = 
1/a_{col} - 1$. The radius $R$ of the lump starts from $R_i$; its 
evolution is then computed by taking into account, besides of the 
varying energy density ($\rho_m$) of the baryon and dark matter 
within $R$, also the energy density and pressure of DE ($\rho_{DE}$ 
and $p_{DE}$, respectively), within $R$ itself. In models involving 
just a cosmological constant, it is $-p_{DE} = \rho_{DE} = {\rm const.}$, 
during the whole process (see, e.g., Lahav et al 1991). In the present 
case, instead, $p_{DE}$, $\rho_{DE}$ and their ratio vary in time,
simply because of their overall evolution with $a$. At turn--around
the radius will be $R_{ta}$. Afterwards, $R$ decreases and should 
vanish at $z_{col}$. The ratio between the energy density
inside $R$ and average yields $\Delta$.

\begin{figure}
\begin{center}
\includegraphics*[width=9cm]{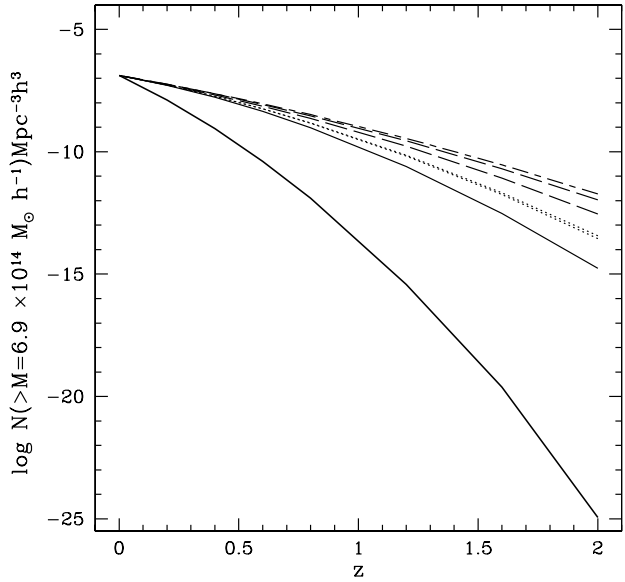}
\end{center}
\caption{The number density of clusters is shown for standard CDM
(lowest curve), OCDM (long--short dashed upper curve),
$\Lambda$CDM (solid intermediate curve),
2 RP (long dashed curves) and 2 SUGRA (dotted curves) models.
The values of $\Lambda/$GeV are $10^2$, $10^6$ for RP and $10^6$, 
$10^{10}$ for SUGRA, starting from the lower curves.
}
\end{figure}

\begin{figure}
\begin{center}
\includegraphics*[width=9cm]{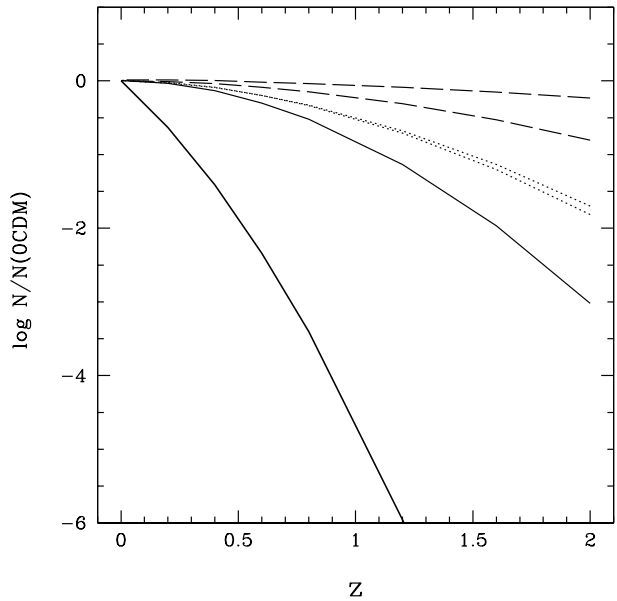}
\end{center}
\caption{The number densities of the previous plot normalized to 
the values in OCDM are shown.
}
\end{figure}

For the sake of example, in Fig.~8 we show $R(t)$ for two RP 
potentials ($\Lambda/{\rm GeV} = 10^2$ and $10^8$) and $\Lambda$CDM. 
SUGRA potentials yield curves intermediate between RP and $\Lambda$CDM
models and just slightly dependent on $\Lambda$. A careful inspection
of the figures shows that, at variance form SCDM and $\Lambda$CDM,
RP models have a slight asymmetry between expansion and recontraction,
due to the evolution of DE density. In Fig.~8, the virialization
radius $R_{vir}$, from which $\Delta_c$ is computed, is also indicated.
In the SCDM case, energy conservation yields $R_{vir}=R_{ta}/2$.
In the presence of DE, requiring virial equilibrium and energy conservation
leads to the cubic equation
$$
x^3 - \left[{ 2+\eta(a_{ta}) \over 2\eta(a_{col}) }\right] x + {1 \over
2 \eta(a_{col})} = 0 ~.
\eqno (2)
$$
Here $x=R_{vir}/R_{ta}$ and
$$
\eta (a) = 2 {1-\Omega_{m} (a) \over \Omega_m (a) (1+\Delta_i)}
\left( R_{ta} \over R_i \right)^3
\left( a_{i} \over a \right)^3
\eqno (4)
$$
($\Omega_m (a)$ is the matter density parameter when the
scale factor is $a$; accordingly, $\Omega_m(1) \equiv \Omega_m$).
Solving such equation yields $x$ values slightly below 0.5$\, $.

\begin{figure}
\begin{center}
\includegraphics*[width=9cm]{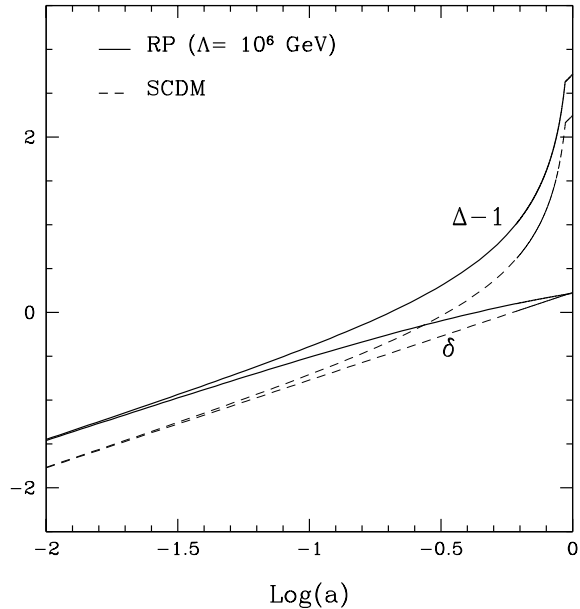}
\end{center}
\caption{Linear and non--linear evolution of density fluctuations.
A models with DE due to a self--interacting scalar field is compared
with SCDM. Similar plots can be given when full recollapse is expected
when $a < 1$. The density contrasts $\Delta_c$, shown in Fig.s 3 and 4,
are $\Omega_m \Delta$.
}
\end{figure}

\begin{figure}
\begin{center}
\includegraphics*[width=9cm]{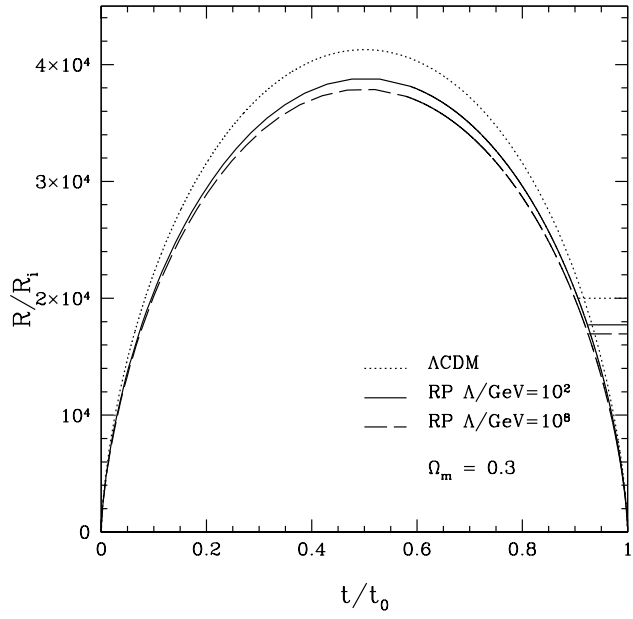}
\end{center}
\caption{Time dependence of the radius of a density fluctuation
in various models; horizontal lines, at the right, indicate the
$R_{vir}/R_i$ attained at $t_o$. Similar plots can be given when
full recollapse is expected at $t < t_0$.
}
\end{figure}

\section{Conclusions}
Observable effects of the nature of DE have been considered
by various authors. For instance, Cooray \& Huterer (1999) discussed
the relation between DE nature and gravitational lensing. 
Previous work on the value of $\delta_c$ was made by Steinhardt, Wang 
\& Zlatev (1999), although explicit outputs were not given. More 
recently, Lokas (2002) considered the behavior of $\Delta_c$ and 
the mass function in DE models with constant $w = -p/\rho$.
This approximated treatment has been pursued in a number of
recent papers and eases computations. Let us however remind that
the value of $w$, when dynamical DE is considered,
varies significantly in the relevant period. In Fig.~9 we report
its variations between $z=0$ and 10, when structures form.
Notice that the rate of $w$ variation is highest in the most critical
redshift interval, between $z=0$ and 1--2. In RP models such variation 
is $\sim 20\, \%$. In SUGRA models it is even greater, as $w$ passes 
from values $\sim 0.8$, at $z=0$, up to values $\sim 0.3$--0.4 in the 
above narrow $z$ interval. Taking constant $w$, therefore, is a
dangerous approximation, whose reliability ought to be carefully
inspected, in different problems.

\begin{figure}
\begin{center}
\includegraphics*[width=9cm]{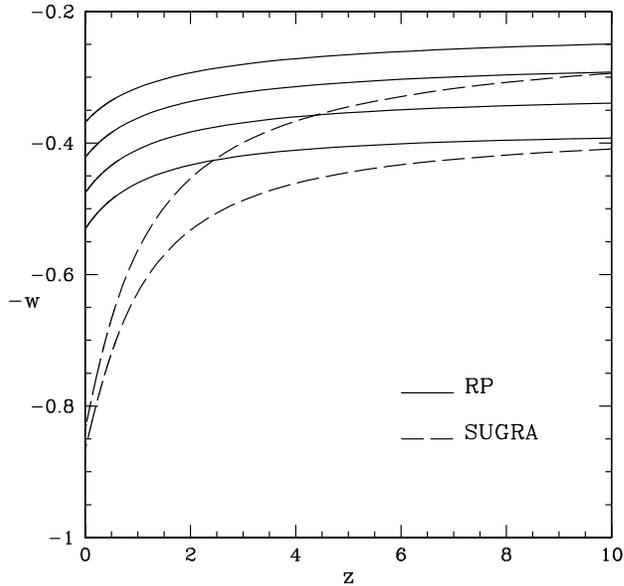}
\end{center}
\caption{Redshift dependence of $w$ for 4 RP ($\Lambda/$GeV =  
$10^2$, $10^4$, $10^6$ and 10$^8$) and 2 SUGRA models ($\Lambda/$GeV =  
$10^2$ and $10^8$). $\Lambda$ values decrease from top to bottom curves.
}
\end{figure}

The main results of this work are that: (i) the shape of the mass function
of clusters, at $z=0$, is only mildly dependent on DE nature.
On the contrary, (ii) the cluster evolution depends on the nature of DE
in a significant way. More in detail, models with RP potentials 
closely approach the evolution expected in open CDM models. 
Only for $\Lambda$ values as low as $\sim 10^2\, $GeV, the expected 
behavior in a RP model is appreciably different from an open 
CDM with the same $\Omega_m$. On the contrary, the evolution of 
SUGRA models is intermediate between open CDM and $\Lambda$CDM models.

Cluster data available within a few years were thought to be able
to discriminate between open CDM and $\Lambda$CDM, on the basis of the
redshift dependence of cluster abundance. If independent data confirm
that we live in a spatially flat world, finding an evolution
closer to open CDM than to $\Lambda$CDM will provide a precise
information on the nature of DE.

As a by--product of the analysis leading to these conclusions,
we also found the dependence of the virialization density contrast
$\Delta_c$ on $\Omega_m$, on the nature of DE, and on redshift $z$.
Such $\Delta_c$ is to be used in various applications, e.g. to
build SO algorithms able to find clusters in n--body simulations
of models with DE.

\end{document}